# Event-based detectors for laser guide star tip-tilt sensing

**Monique Cockram*** **and Noelia Martinez Rey**
Australian National University, Research School of Astronomy and Astrophysics, Advanced Instrumentation and Technology Centre, Canberra, Australian Capital Territory, Australia

**ABSTRACT.** Event-based sensors detect only changes in brightness across a scene, with each pixel producing an asynchronous stream of spatial-temporal data, rather than recording frames of overall illumination such as a traditional frame-based sensor. This is advantageous for implementing into a wavefront sensor, which benefits from high temporal resolution and high dynamic range. The determination of tip-tilt in particular is still a problem in laser guide star (LGS) adaptive optics as there are no current technological capabilities to measure it. We characterized the behavior of an event-based sensor in the context of tip-tilt sensing, investigating if the high temporal resolution of the event streams could address these challenges. Different conditions of tip-tilt and background illumination levels are explored and found to be a strong contender for tip-tilt sensing with LGSs.

## 1 Introduction

Sodium laser guide stars (LGSs) are artificial stars generated in the upper atmosphere by exciting sodium atoms maintained by the deposition of meteoritic ablation particles at altitudes between 80 and 100 km above the Earth's surface. The laser source that excites these sodium atoms is propagated from the ground up to the mesosphere, generating the LGS and resulting in photons traveling downward toward the telescope's collecting aperture. This dual propagation is one of the main differences between the light coming from an LGS and from a Natural Guide Star, a source located at infinity whose photons only travel downward. For this reason, conventional adaptive optics (AO) techniques using an LGS do not enable the measurement of tip-tilt aberrations, thus necessitating their combination with natural guide stars.

Most AO systems create the LGS using a bistatic configuration where the laser is propagated by a separate launching facility, and the LGS photons are captured by the main telescope.[1] In this configuration, the LGS wavefront sensor measures both the uplink tip-tilt (caused by turbulence affecting the laser upon exiting the launch telescope) and the global downlink tip-tilt. Decoupling each contribution has not yet been proven possible. This issue is known as the LGS tip-tilt indetermination problem.[2]

AO, although developed for astronomy, is also important for free-space optical communications with satellites.[3] The optical signal needs to be pre-compensated for atmospheric turbulence to be received by the satellite with minimal error.[4] This expands the use cases of AO into daytime conditions, placing further limitations on tip-tilt sensing.[5]

*Address all correspondence to Monique Cockram, monique.cockram@anu.edu.au

### 1.1 Point Source Laser Guide Star and the Time-Delay Method

A different approach in LGS-AO utilizes monostatic configuration,[6] such that the laser is propagated using the same telescope that collects the LGS photons. In a monostatic launch, the LGS tip-tilt indetermination arises from the fact that both guide star laser propagation paths (upward and downward) travel through almost the same volume of the atmosphere. As a result, any tip or tilt in the beam path caused by the atmospheric turbulence is canceled out almost entirely by its return propagation.[7]

Even in this configuration, the LGS tip-tilt indetermination problem remains unsolved. However, the fact that the uplink and downlink tip-tilt do not fully cancel each other out, leaving differential tip-tilt on the order of milliarcseconds, is a promising avenue to explore. The time-delay method was first proposed by Ragazzoni[8] in the 1990s and has not yet been validated due to a lack of appropriate technology. The CaNaPy system[9,10] is likewise aiming to validate this method, using a Pyramid wavefront sensor in a monostatic configuration.

The time-delay method, which involves measuring the small evolution in the tilt signal that is not canceled out by the light's return propagation through the atmosphere, is demonstrated in Fig. 1. The small remaining differential tip-tilt, denoted by the dashed line, is measured to determine tip-tilt from the LGS. The time-delay method is affected by the turbulence sampling on the LGS and the accumulated integration time of a series of measurements. The maximum integration time before the error exceeds the diffraction limit is[8]

$$\Delta t \approx \left(\frac{D_p}{r_0}\right)^2 \frac{P}{P_0} \frac{\tau^2}{t^*}, \tag{1}$$

where $D_p$ is the telescope diameter, $r_0$ is Fried's parameter, $P/P_0$ is the fraction of minimum laser power for full correction used, $\tau$ is the time delay of the laser propagation to the sodium layer of the atmosphere, and $t^*$ is the sampling time of the LGS. This maximum integration time provides an indication of how often the drift on the LGS-retrieved absolute tip-tilt needs to be corrected by the absolute measured tip-tilt on an natural guide star (NGS).

An example case of a 1-m telescope observing through an atmosphere of coherence length $r_0 = 10$ cm and $P/P_0 = 1$ is herein illustrated. The time delay to the sodium layer at 90 km altitude at the zenith is assumed to be $\tau = 0.6$ ms. In these conditions, an LGS wavefront sensor with sampling time $t^* = 1$ ms (scenario A) would need to measure tip-tilt from an NGS every 36 ms to correct the drift and remain diffraction-limited. For a sampling time $t^* = 0.5$ ms (scenario B), the maximum time between NGS corrections is doubled to 72 ms. As a result of these NGS corrections, the sky coverage is increased by a factor $\eta$,[8] as described by the following:

$$\eta \approx \left(\frac{\Delta t}{t^*}\right)^{\frac{3}{2}}. \tag{2}$$

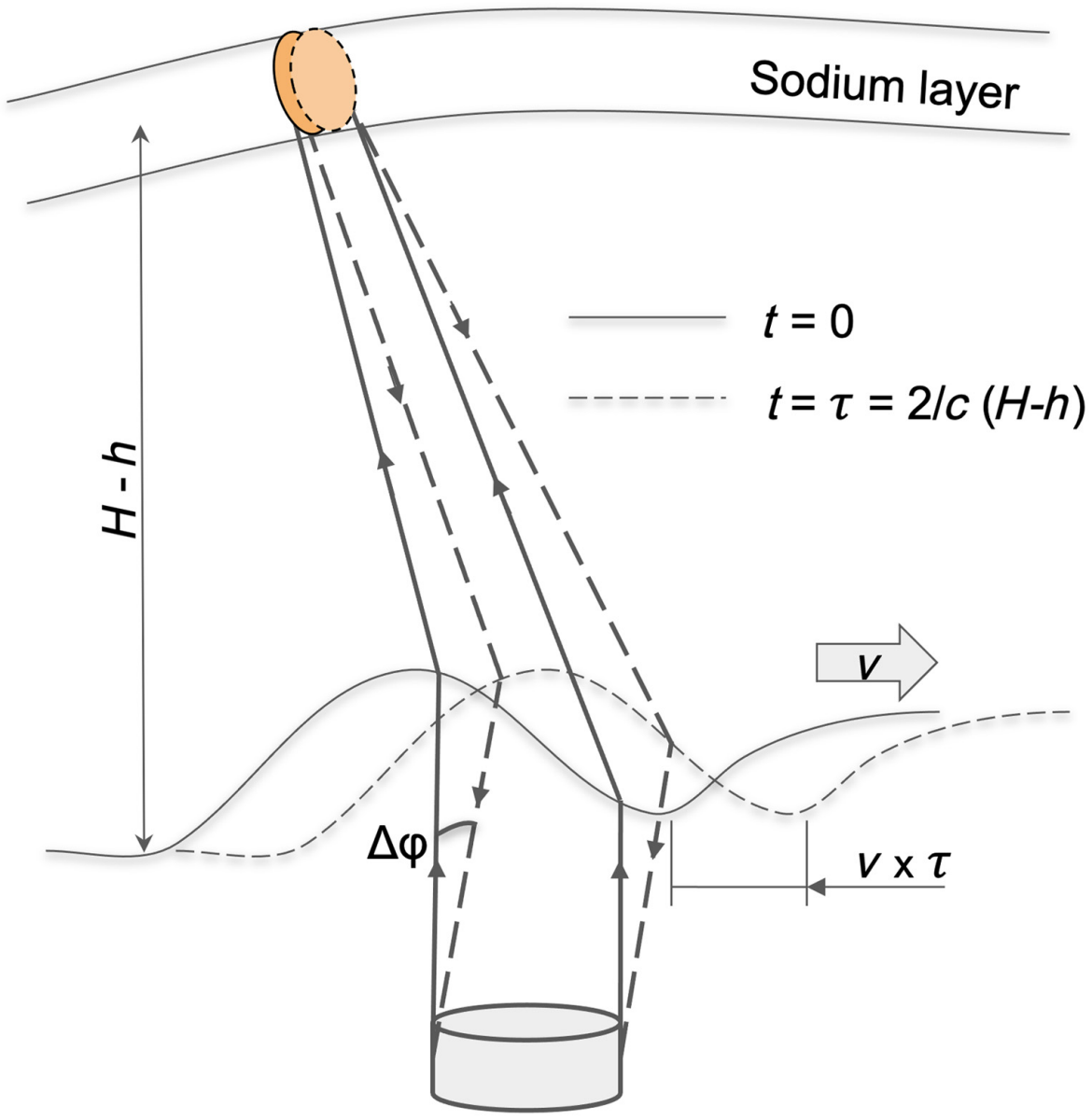

**Fig. 1** Time-delay method. Adapted with permission from Ragazzoni.[8] Here, $H$ is the height of the sodium layer in the atmosphere; $h$ is the ground layer of the turbulent atmosphere; $v$ is the wind speed; c is the speed of light; $t$ is the propagation time; $\tau$ is the time delay between the upward and downward propagation of the LGS; and $\Delta\varphi$ is the differential tip-tilt due to the time delay measured by the wavefront sensor (WFS).

In the example scenarios, this results in a sky coverage increase factor of $\eta = 215$ in scenario A and $1.73 \times 10^3$ in scenario B. This demonstrates how doubling the tip-tilt sensing speed results in an increase of $\eta$ by a factor of 8 and places a limit on the integration time of the LGS wavefront sensor for larger sky coverage. Although shorter integration times limit sky coverage when using an NGS wavefront sensor (brighter NGS required), the reduced reliance on an NGS using the time-delay method means that an appropriate NGS nearby is required even less frequently for even higher speed corrections.

The time-delay method has not yet been demonstrated due to the characteristics and capabilities of wavefront sensors that use frame-based detectors. With these detectors, the real-time analysis of the wavefront within the AO loop has a finite processing speed based on a set computing power, which restricts the feasible speed and resolution of each of the measurements of the atmosphere. The very small differential tip-tilt that the time-delay method aims to retrieve is likely outside these performance restrictions and most likely within the noise floor level of such detectors. To measure and correct for tip-tilt at the diffraction limit, one needs to measure the absolute tip-tilt with higher accuracy. Considering the diffraction limit of a 1-m telescope, the absolute tip-tilt measurement accuracy should be $>150$ milliarcseconds. However, the differential tip-tilt is typically one order of magnitude smaller than the absolute tip-tilt. To achieve diffraction-limited performance in the absolute tip-tilt derived from differential tip-tilt, the latter needs to be measured with milliarcsecond levels of accuracy.

This research aims to study the feasibility of a new type of detector technology, event-based sensors, with the potential to detect and measure differential tip-tilt on LGSs using the time-delay method. Event-based sensors detect asynchronous changes in the scene, removing the restriction of data to pre-determined frame rates and increasing temporal resolution. By increasing the speed of LGS corrections using the time-delay method, the sky coverage is significantly increased, as in Eqs. (1) and (2). The asynchronous data output is significantly smaller than that of a frame-based detector, reducing the computational power required for real-time sensing. This sensing regime could be advantageous for detecting small tip-tilt movements as the data output should only contain these changes and not a full-frame image. However, as there is no set integration time, a meaningful measurement will take place after a sufficient number of events have been triggered. Triggering of events is directly related to the tip-tilt amplitude as the LGS differential movement will cause a change in illumination in the event-based camera field of view. As a result, the optical configuration of these detectors (plate scale) needs to be dimensioned according to small tip-tilt amplitudes. In addition, any changes due to fluctuations in noise could saturate the data more significantly than a frame-based detector. As a result, these noise events need to be carefully considered and masked. This paper will discuss these characteristics and their impact on LGS tip-tilt sensing.

### 1.2 Event-Based Sensors

Where traditional frame-based sensors register the overall illumination during a fixed exposure time, event-based sensors detect only local brightness changes in the scene.[11] These sensors produce an asynchronous stream of spatial-temporal events data, outputting single-pixel information as each experiences a change in illumination. The output data contain a stream of events labeled with spatial location on the sensor, timestamp of the event, and polarity of the change in logarithm of illumination.

Figure 2 shows that the sensor is composed of a photoreceptor and an amplifier that outputs the logarithm of the signal, followed by a differencing circuit that amplifies any changes in illumination for each pixel by comparing it with the stored memory of the illumination of the previously triggered event. Subsequently, a comparator compares those changes in illumination with a set threshold value and determines the polarity of the event (i.e., an increase or decrease in illumination). For this sensing principle, the illumination level of the previous event is captured in the circuit to use as a comparison in the differencing circuit.[13–18] A faster illumination change will generate more events in a particular timeframe than a slower illumination change, as visualized in Fig. 2. Background noise and fluctuations can be filtered by appropriately choosing a threshold for what magnitude of illumination change is considered an event. In addition, any constant background is not sensed by the detector in a theoretical noise-free environment due to its operational principle of only detecting the changes in the light within the field of view. The appropriate contrast threshold levels vary between individual sensors and with the conditions of imaging.

**Fig. 2** Operational diagrams of an event-based sensor. Adapted with permission from Ref. 12. (a) Circuit diagram of an event-based sensor. (b) Demonstration of how changes in illumination are represented by ON and OFF events, triggered as the signal crosses the horizontal event threshold (dashed lines).

The asynchronous nature of event-based detectors means pixel read-out only occurs when and where an event is triggered. This significantly changes the read-out noise as the full pixel array is not read out at every instance. This operational principle also benefits a high temporal resolution as the detector is not restricted to a constant frame rate. Instead, the pixels are read out asynchronously at the time when a change triggers an event. The individual pixel event stream is of a much lower bandwidth output than a classical detector's full-frame output, resulting in advantages for implementation into real-time sensing applications and high-speed calculations. However, the clear difference in sensing method and data output from that of a classical detector creates challenges for seamlessly integrating an event-based detector into existing instruments. Any control loops or analysis algorithms would require adaptation to accept this unique data format as all existing instruments would be designed for the same data output of traditional frame-based sensors.

The particular advantages of these detectors include a high temporal resolution, ideal for sensing fast atmospheric changes. Their high dynamic range allows for their application in a range of illumination levels (daytime wavefront sensing). Other applications have been reviewed by Gallego et al.[11] in 2020, including initial applications in self-driving cars,[19] robotics,[20] and object tracking.[21] More research has begun in astronomical contexts in recent years, such as Ralph et al.[22] used an event-based detector for astrometry, and Afshar et al.[23] developed an event-based dataset of space and astronomy images. Cohen et al.[24,25] used an event-based detector to track satellites in different orbits around Earth. In addition, Krüger and Kamiński[26] demonstrated an accuracy in the astrometry of satellites of 1.5 arcseconds. Particularly, relevant to this work were Kong et al.,[12] Grose et al.,[27] and Ziemann et al.,[28] who implemented a Shack-Hartmann wavefront sensor with an event-based detector.

These applications and the specifications of event-based detectors indicate the potential advantages of event-based detectors to a wide range of wavefront sensing applications. Specifically, this research focuses on the application of event-based wavefront sensors to the retrieval of tip-tilt using LGSs.

## 2 Operation Principle

The model of an event-based detector is such that an event is generated when the logarithm of intensity difference at a given time is greater than or equal to the contrast threshold. An event is generated if the logarithm of the intensity from the photodetector, $L \equiv \log I$, at pixel $\vec{x}_k = (x_k, y_k)$ at a time $t_k$ satisfies[11]

$$\Delta L(\vec{x}_k, t_k) \equiv L(\vec{x}_k, t_k) - L(\vec{x}_k, t_k - \Delta t_k), \tag{3}$$

such that

$$\Delta L(\vec{x}_k, t_k) = p_k C, \tag{4}$$

where $C$ is the contrast threshold, and $p_k \in \{-1, +1\}$ is the polarity of the change (i.e., increase or decrease in illumination).

The circuitry of the event-based detector takes the logarithm of the light intensity of the photodetector, notably before the differencing circuit. This means that the differencing process that enables the event-based sensor's immunity to sky background is not the very first process to happen. As a result, even a constant background does have an effect on the response of the detector. This is outlined in Eq. (5), where two signals $I_1$ and $I_2$ with a background level $b$ will trigger an event if the difference is sufficient. It is clear how the background $b$ still remains a factor but would be canceled out with simple subtraction of the two signals if the logarithm was not applied

$$\text{difference} = \log(I_1 + b) - \log(I_2 + b), \tag{5}$$

$$= \log\left(\frac{I_1 + b}{I_2 + b}\right). \tag{6}$$

This is shown in Fig. 3, which plots the theoretical response of an event-based detector [Figs. 3(b)–3(d)] to a sinusoidal signal [Fig. 3(a)]. From Fig. 3(b), it is clear that the regions with the lower laser signal strength generate many more events when the background illumination is also low. As the background level increases, this asymmetrical response to different laser source intensities is dampened, eventually returning to the same sinusoidal pattern of intensity as the signal input into the detector. The dotted lines that represent a threshold value of the event-based sensor can be used to visualize how each curve triggers an event (i.e., crosses to the next integer multiple of threshold value) at different rates. As a result, lower background illumination levels result in more events being triggered, particularly for low signal strength, as described in Sec. 2.1. If the contrast threshold value [Eq. (4)] is increased, the separation of the dotted lines would increase, and the intensity curves would cross these thresholds less often for a similar change in illumination.

Figure 3(c) shows a similar effect. Here, the behavior of the sensor remains relatively sinusoidal until the fluctuation in laser power exceeds the level of the background illumination. In these conditions (low-level power), the signal triggers many more events. This is also observable in Fig. 3(d); here, the increased laser power results in the intensity curve crossing the threshold lines more times.

These theoretical representations of the event-based sensor's response to different conditions indicate that its performance in a dynamic scene, such as this sinusoidal signal, is best when the background illumination is greater than the signal illumination level. These plots show that the relative intensity between background and signal is an important factor to consider.

This behavior due to the logarithmic intensity signal results in an uneven response of the detector, making it difficult to choose an appropriate threshold value across laser powers. If a signal is being detected only in a small section of the dynamic range of the detector, this effect is less significant. However, making use of the full dynamic range of the detector in a single measurement results in an uneven response, as illustrated in Sec. 2.1.

### 2.1 Understanding Noise in Event-Based Sensors

The analysis of the signal-to-noise ratio (SNR) of an event-based camera differs from that of a classical detector. In a classical detector, a simple shot noise limited SNR is $\frac{I}{\sqrt{I}}$, where the signal $I$ is the detected intensity of light. The initial amplifier in the event-based detector circuitry outputs the logarithm of the signal, which is a significant difference from that of a classical detector. An event-based detector requires consideration of what makes a detectable event, that is, if the ratio in Eq. (6) is above the threshold to trigger an event. In this case, the signal used in the SNR calculation is $\log(I)$.

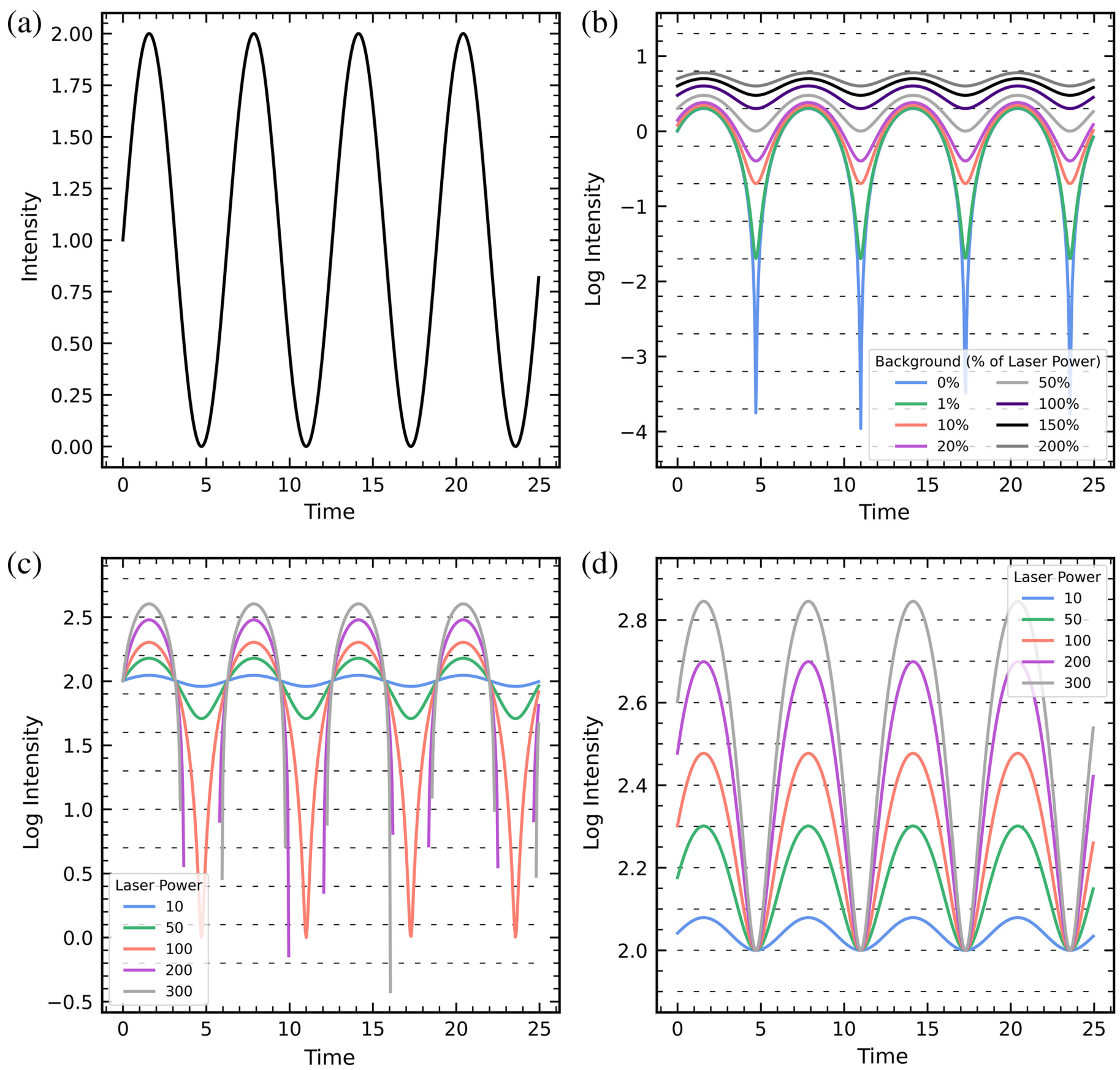

**Fig. 3** Theoretical plots of the response of the event-based sensor to different input intensities and background levels. The dashed lines are used to represent a threshold for triggering an event, where the spaces between the lines are set to the threshold value. Plots are depicted as intensity in arbitrary units. (a) Sinusoidal intensity input signal. (b) Increasing background as a percentage of (constant) laser power. (c) Increasing the magnitude of change in laser power. Background level (100 units) and average laser power are kept constant. The legend shows the amplitude of the sinusoidal laser signal. (d) Increasing the magnitude of change in laser power as well as the average laser power. The background level is kept constant at 100 units. The legend shows the multiplication factor applied to both the average laser power and the amplitude of the change in laser power.

By analyzing the contribution of different noise sources, one realizes how event-based detectors have the potential to dominate over frame-based detectors in given applications. Readout noise is significantly smaller for an event-based detector because only pixels that have triggered events are read out, whereas a frame-based detector regularly reads out the full frame of pixels. Absolute illumination (signal + background) of an event-based detector significantly impacts its noise output. The spread of background activity shows this characteristic, where low-brightness scenes are dominated by signal-dependent noise sources such as shot noise.[29] Section 3.2.2 demonstrates experimentally how low illumination can result in up to $10^3$ times the event rate of shot noise than high illumination scenes. In an event-based sensor, shot noise triggers events significantly more when the illumination is low due to the capacitor in the pixel circuitry not being saturated. In brighter illumination conditions, the detector response to noise is instead dominated by leak events, caused by the decay of the stored analog memory. The values of previous events stored as voltage on a capacitor are required by the detector's operating principle. This decays linearly with time, and results in the drifting of the previous illumination value toward generating another ON event.[30] As a result, leak noise is dominated by ON events. The rate of this decay increases with absolute input illumination, causing background pixels to be less sensitive to shot noise and creating overall less background noise for scenes with higher illumination levels. This noise source is deterministic, rather than random, as its rate can be determined by characterizing the drift of the capacitor.

The magnitude of background activity can be minimized by changing the event threshold bias [contrast threshold, Eq. (4)] and applying background activity filters[23,31,32] to the event stream.

Section 3.2.2 outlines the experimental characterization of event-based sensor noise.

## 3 Characterization of an Event-Based Sensor

### 3.1 Experimental Setup

A CenturyArks SilkyEvCam VGA event-based camera (containing the Prophesee PPS3MVCD sensor with pixel pitch 15$\mu$m) has been tested in a dedicated test bench at the Advanced Instrumentation and Technology Centre in the Australian National University (ANU). To introduce a set of known aberrations, the Thorlabs DMP40 deformable mirror was used as a tip-tilt mirror in combination with a 589-nm low-power laser source. The deformable mirror (DM) was used to only actuate tilt; a range of angle amplitudes, as well as several change rates, were tested to verify the capabilities of the event-based detector and determine its dynamic range. A lens with a focal length of 200 mm was included to set plate scales of 15 arcsec/pixel. The experiments were conducted with background illumination from ambient laboratory lights, along with additional artificial illumination placed next to the focusing lens. This setup resulted in illumination levels that increased by approximately a factor of 2 in measured brightness at each step. This setup allowed for the following variables: laser power, background level, tilt frequency, and tilt amplitude.

The tilt was calculated using an average position tracker algorithm[12] on the event-based data, recording the changes in the movement of the laser spot on the detector. This algorithm, outlined in Eq. (7), updated the average position $(\hat{x}_{\mathrm{ti}}, \hat{y}_{\mathrm{ti}})$ at the event timestamp every time an increase (or ON) event $e_i = [x_i, y_i, t_i]$ was generated by the detector. The weighting parameter $m$ can be used to control how quickly the position changes as each new event is acquired. If $m = 0$, the previous position estimate does not contribute at all to the current position. As $m$ increases toward 1, the position calculation has greater inertia toward its previous estimate. The best value of $m$ for these data was determined to be 0.8 and was optimized empirically based on the drift of the position estimate with respect to the true spot position

$$\hat{x}_{\mathrm{ti}} = m\hat{x}_{\mathrm{ti-1}} + (1 - m)x_i, \tag{7}$$

$$\hat{y}_{\mathrm{ti}} = m\hat{y}_{\mathrm{ti-1}} + (1 - m)y_i. \tag{8}$$

The performance of the event-based detector can be adjusted using parameters called biases. This setting controls the level of illumination change that triggers an ON (bias_diff_on) or OFF (bias_diff_off) event, relative to the bias_diff parameter. Low pass (bias_fo) and high pass (bias_hpf) filters can further control what illumination changes trigger events. The refractory period, or the time period that each pixel is deactivated after each event, can be adjusted using bias_refr. When referring to contrast threshold values, this is the difference bias_diff_on −bias_diff.

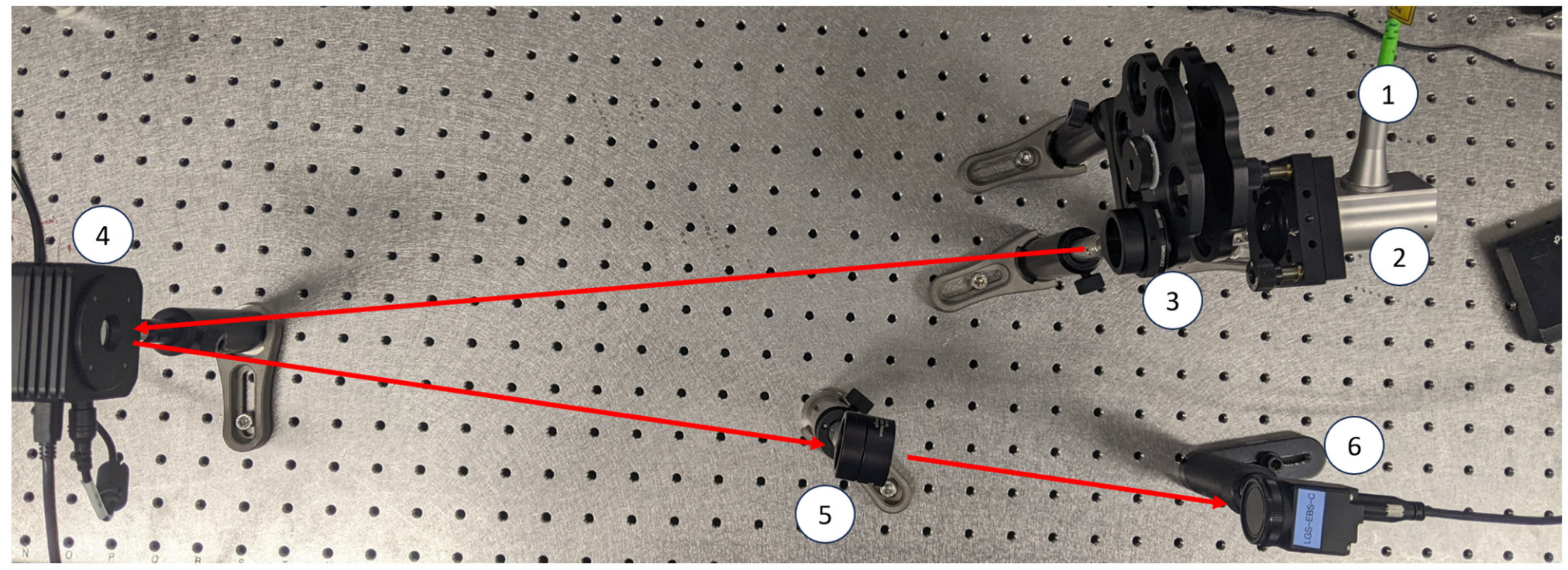

**Fig. 4** Laboratory test bench for the characterization of the event-based detector, labeled as in Table 1.

**Table 1** Components used in the testing, as in Fig. 4.

| Component | Part number | Label (as in Fig. 4) |
|---|---|---|
| Laser (589 nm) | MSL-U-589-5 mW | 1 |
| Collimator | Thorlabs RC12APC-P01 | 2 |
| Iris | Thorlabs SM1D12CZ | 3 |
| DM | Thorlabs DMP40 | 4 |
| Lens | 200 mm: Thorlabs AC254-200-A-ML | 5 |
| Event-based detector | CenturyArks SilkyEvCam (VGA) | 6 |

**Table 2** Bias settings for the event-based detector.

| Type | Value (mV) |
|---|---|
| bias_diff | 299 |
| bias_diff_on | 384 |
| bias_diff_off | 221 |
| bias_fo | 1560 |
| bias_hpf | 1448 |
| bias_refr | 1500 |

The values of the bias settings (unless otherwise stated in Sec. 3.2) for the event-based detector are outlined in Table 2. Note that only bias_fo was changed from the default values.

### 3.2 Results

The event-based detector experiments investigated its response under different illumination conditions. Its accuracy of tip-tilt measurements for a range of background and signal levels is explored in Sec. 3.2.1, and the detector's noise characterization is in Sec. 3.2.2.

#### 3.2.1 *Detector accuracy across illumination levels*

The following experiment aimed to verify the theoretical behavior described in Sec. 2. Changing the background illumination level of the set-up was achieved by introducing external light sources, placed next to the focusing lens. Concurrently, the laser power was attenuated using neutral density filters to also investigate the effect of variable laser intensity. The behavior of the detector was thereby analyzed for a range of illumination levels for both signal and background. The tilt experimental and analysis procedure as described in Sec. 3.1 was used; in this case, the key information was the number of events recorded. The tilt of 200 arcseconds with a frequency of 200 Hz was measured, with 400 oscillations in each run of measurement. The laser power ranged between 0.005 and 0.05 mW using neutral density filters.

Figure 5 shows the results of this experiment. It is clear that measurements made under higher background levels output fewer events than the equivalent test under lower background levels. This agrees with the operational principle in Fig. 3(b), which shows how the detector response became more uniform across the laser power levels as the background level increased. Also consistent with the analysis of the operational principle, the lower background levels produced more events across all laser powers. The results were indeed comparable to the logarithmic intensity plots, which show the detector response curve crossing the marked event threshold (and thus producing an event) more times for lower background levels.

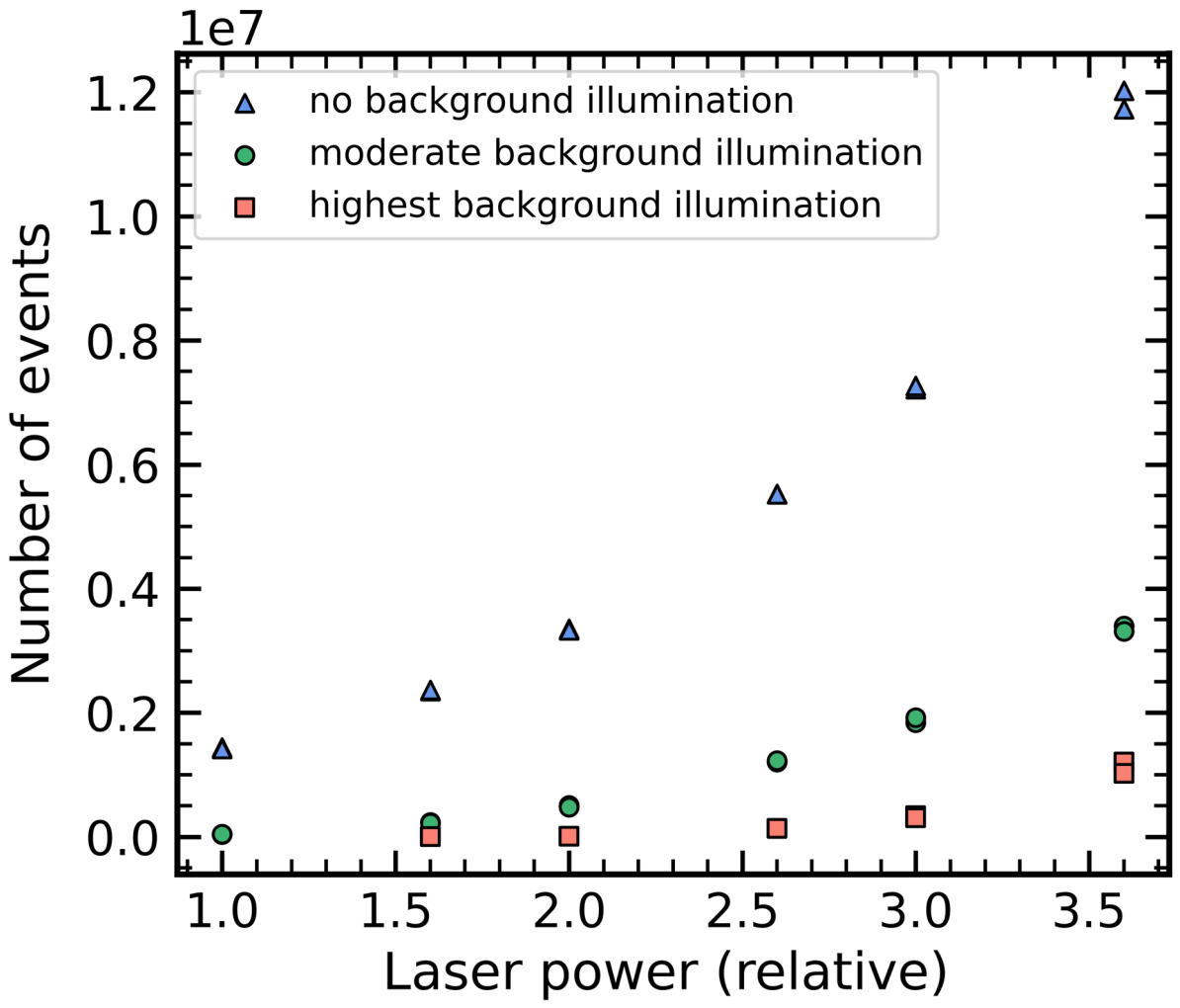

**Fig. 5** Number of events registered during a run of tilt measurements, showing the different number of events recorded as the laser power is attenuated for three background illumination levels. Laser power was attenuated using neutral density (ND) filters, and different backgrounds were introduced using external light sources. No background illumination is a dark room, moderate illumination is overhead ambient lighting, and the highest background illumination includes the addition of a torch adding light to the detector.

Furthermore, the behavior of the detector when the background is kept constant was also verified in each of the three experimental datasets. As laser power increased, so did the number of events recorded. This is comparable to the theoretical plots in Figs. 3(c) and 3(d).

An additional experiment was undertaken to investigate how the threshold values affect the detector response to these different conditions. This aimed to test the detector's capability to have the same response even for different background illumination levels by appropriately tuning the threshold values. The response of the detector to the same tilt signal measured at different threshold values was recorded. This was repeated for two background illumination levels. An example of these tests can be found in Fig. 6, where the conditions of a high background with a threshold value of 50 match the detector response approximately with the moderate background level with a threshold value of 100. This difference in the threshold value of 50 was consistent across multiple tests for this pair of background levels. Conditions could also be matched across three additional background illumination levels, as indicated by the red arrows in Fig. 7.

This ability to match conditions demonstrates that immunity to a constant background can be achieved with the event-based detector despite the background level affecting the response of the detector.

In addition, these experimental data demonstrated that the threshold values have a less significant effect on the response of the event-based detector in higher background illumination conditions. Figure 7 shows how the same range of threshold values had less of an effect on the number of events recorded for conditions where the background is higher. Where there was no background illumination, even small changes in the threshold value significantly changed the number of events recorded. This demonstrates that the response to the chosen threshold values also depends on the background illumination level.

### 3.2.2 *Noise characterization*

The background activity of the event-based sensor was recorded to characterize the types of noise under different lighting conditions in the laboratory. The ambient light conditions were either dark (windowless laboratory space) or bright (laboratory space with ceiling lights) to simulate night and daytime conditions, respectively. An additional condition was that of the laser, which was either on or off; however, no tip-tilt movement was introduced for these measurements. A further test involved covering the sensor with its cap.

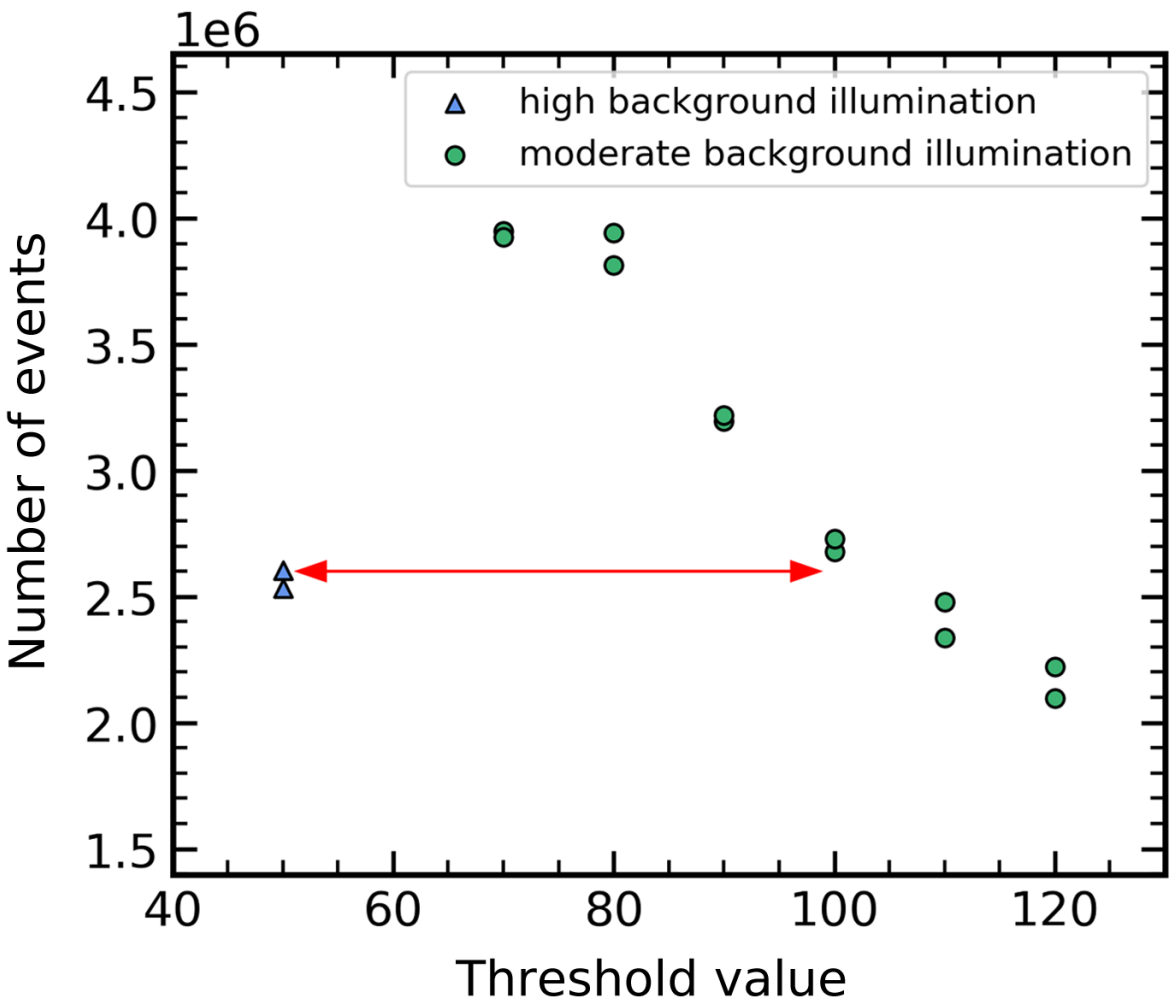

**Fig. 6** Number of events registered during a run of tilt measurements. The number of events recorded for each background illumination level can be made equivalent by choosing an appropriate threshold value. The red horizontal arrow indicates matched conditions between the high and moderate background conditions. Different backgrounds were introduced using external light sources, and the threshold value of the event-based detector was set to a range of values. The threshold values refer to bias_diff_on−bias_diff. No background illumination is a dark room, moderate illumination is overhead ambient lighting, and the highest background illumination includes the addition of a torch adding light to the detector.

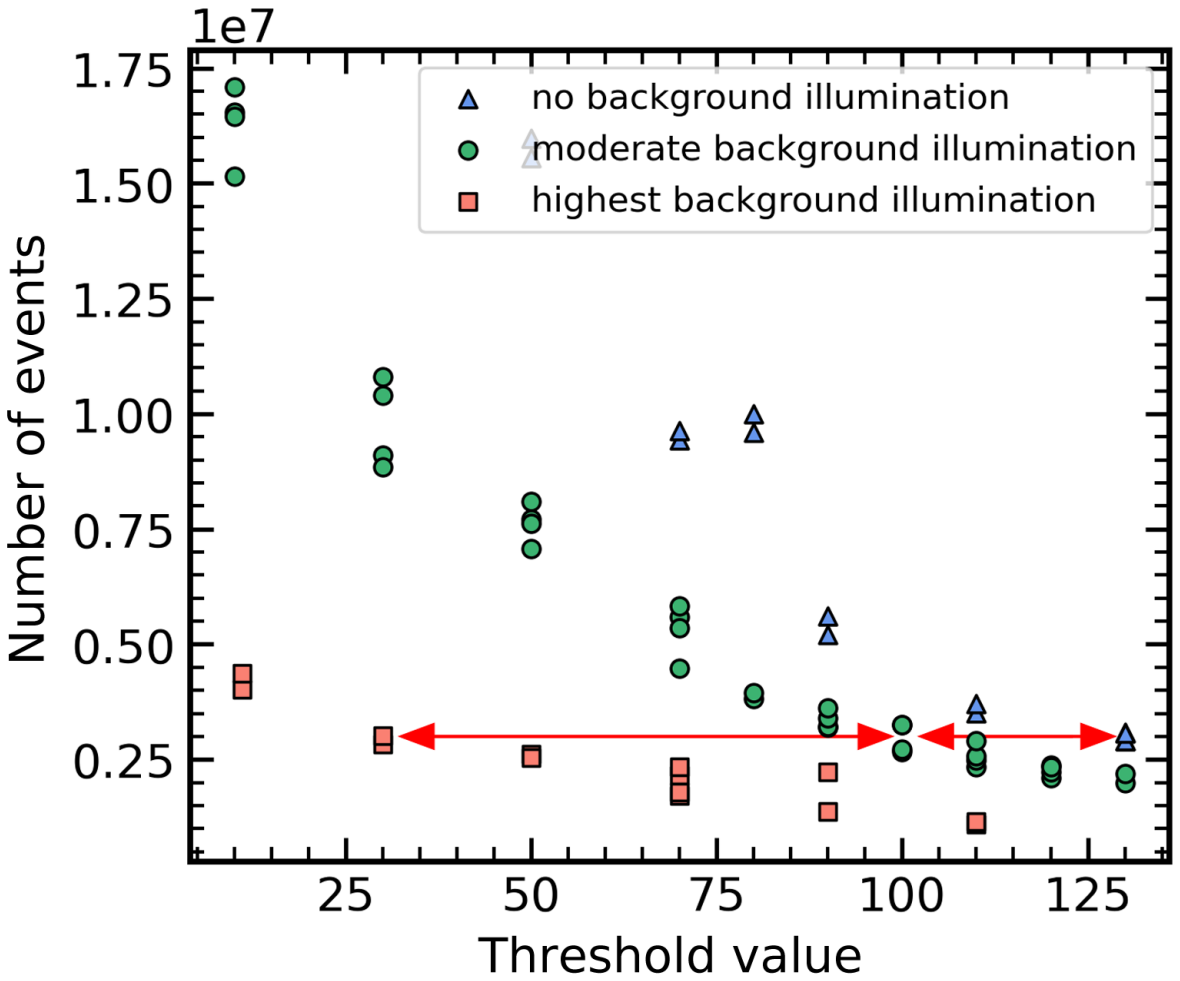

**Fig. 7** Number of events registered during a run of tilt measurements, showing how conditions can be matched across multiple different background illumination levels using threshold values. More than one combination results in matching conditions, one of which is demonstrated by the red horizontal arrows between three different conditions of background illumination. Different backgrounds were introduced using external light sources, and the threshold value of the event-based detector was set to a range of values. No background illumination is a dark room, moderate illumination is overhead ambient lighting, and the highest background illumination includes the addition of a torch adding light to the detector.

The noise measurement in Fig. 8(a) (laser on and background illumination) results in a very high percentage of ON events (positive illumination change) of 99%, specifically at the pixels where the laser spot is detected. A typical scene of average brightness has approximately equal numbers of ON and OFF events. These increased events correspond to the leak events discussed in Sec. 2.1 caused by very high illumination levels. Outside of this area, the background events are minimal as the leak events are reduced, and the background light minimizes the shot noise. The event rate of this noise was measured to be 0.037 events/pixel/s over the area of the detector. The conditions in Fig. 8(c) (laser on and no background illumination) are the least optimal, with significant shot noise from the dark background conditions as well as leak events occurring where the laser hits the detector. This resulted in an event rate of 4.1 events/pixel/s over the whole detector, significantly higher than in Fig. 8(a). The ring of low noise around the laser spot is presumably due to the dispersion of the laser causing faint light to minimize both shot noise and leak events over those pixels.

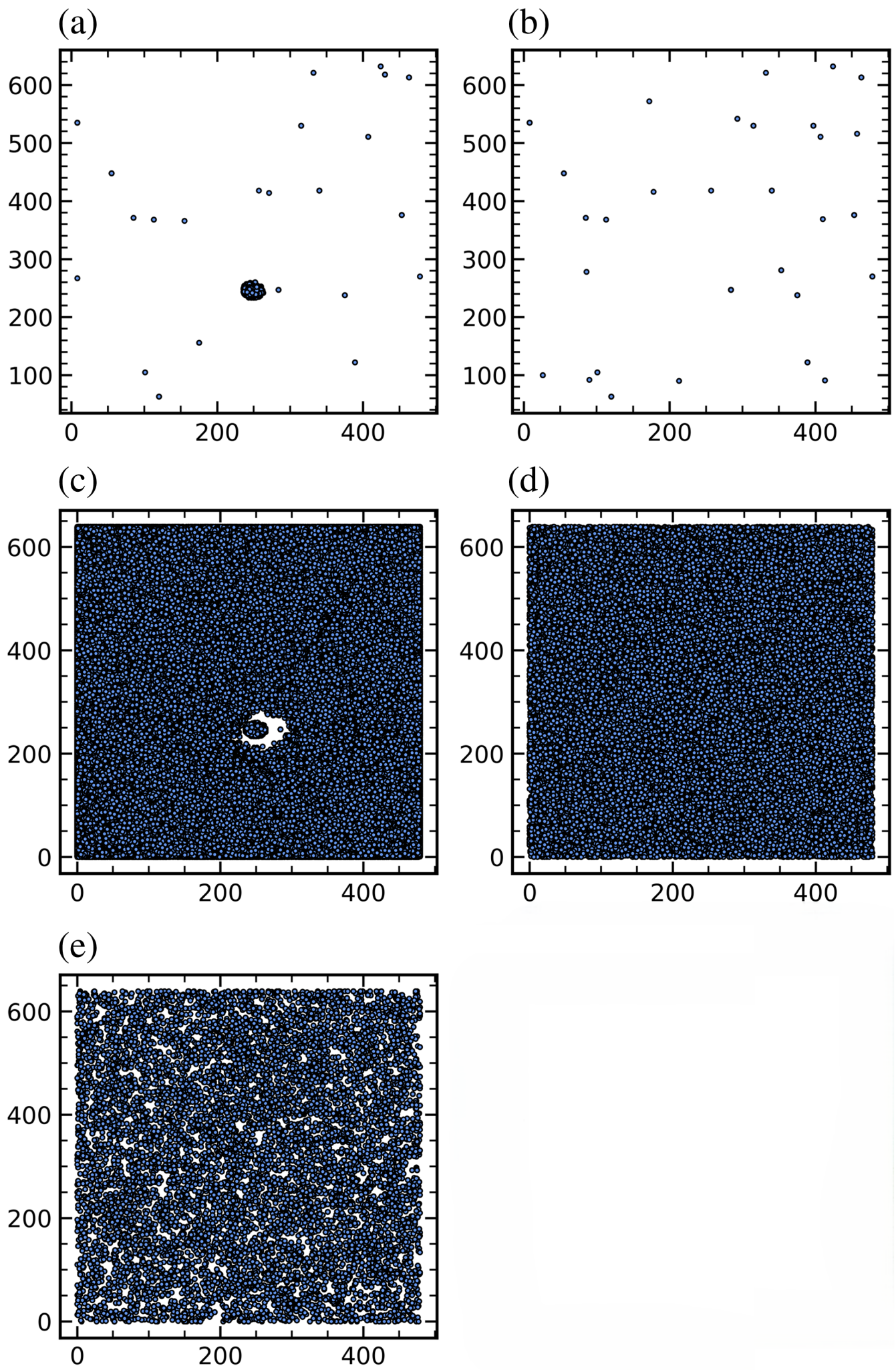

**Fig. 8** Noise only measurements taken with the event-based detector for different combinations of ambient/background light and the presence of stationary laser signal. These frames represent 10-s accumulations of events. The detector configuration remains the same between each image. Dark background refers to a dark room and a bright background is created using overhead ambient lighting. (a) Bright background, laser on. (b) Bright background, laser off. (c) Dark background, laser on. (d) Dark background, laser off. (e) Detector cap on.

For conditions where there is no laser causing leak events [Figs. 8(b), 8(d), and 8(e)], the noise level radically improves in the presence of background illumination. Figure 8(b) demonstrates that bright conditions with no laser illumination that could cause significant leak events have very low noise levels, measured at a rate of $4.8 \times 10^{-5}$ events/pixel/s across the whole area of the detector. However, once the bright background light is removed (ambient light off), shot noise begins to dominate. This is shown in Figs. 8(d) and 8(e), where the noise events are distributed over the entire area. The histogram of distances between successive noise events in Fig. 9 clearly demonstrates its normal distribution. The event rate of this noise is 0.067 events/pixel/s and 0.039 events/pixel/s for Figs. 8(d) and 8(e), respectively.

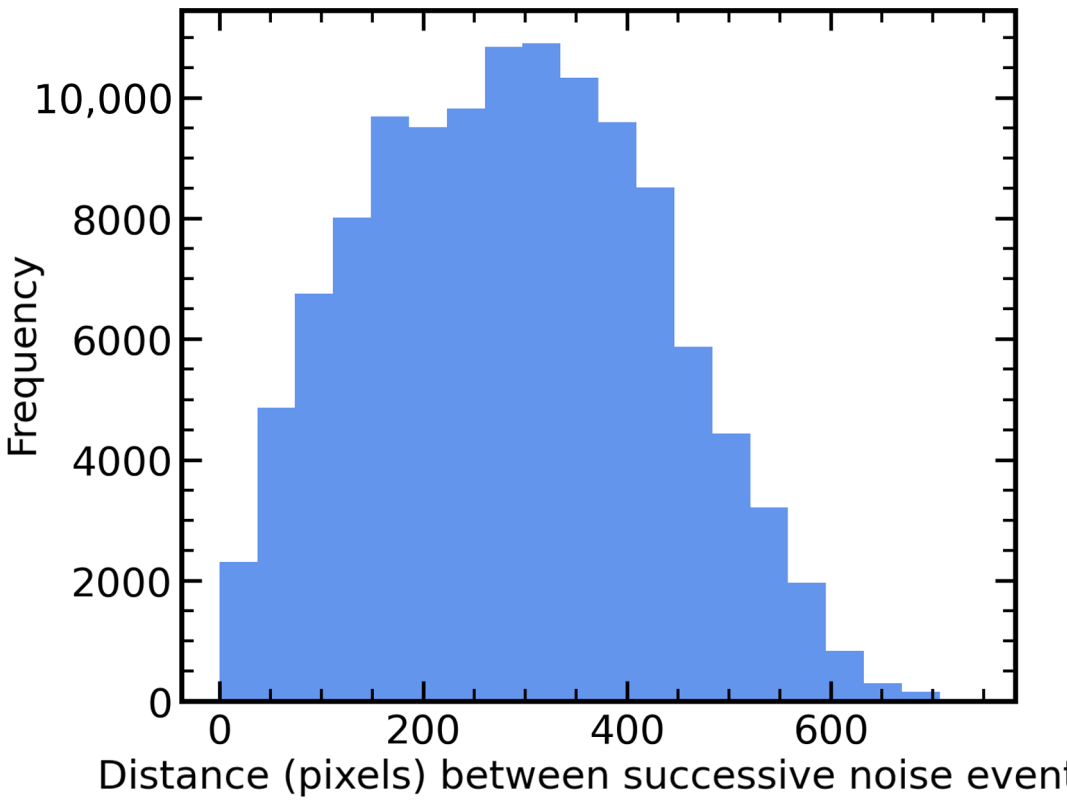

**Fig. 9** Histogram of the pixel distance between noise events shown in Fig. 8(e). The log-normal shape indicates the random distribution of shot noise.

In bright conditions, the capacitors in the event-based sensor's circuit are saturated by the background illumination and thus are unable to detect and trigger events due to shot noise, and the output only shows noise in the form of limited leak events. Where the brightness conditions are lower, such saturation does not occur, and shot noise triggers events across all the pixels.

This relationship between background noise activity and the illumination of the event-based detector pixels is in agreement with Graça et al.[30]

The noise characterization of event-based detectors is further discussed in the context of LGS tip-tilt sensing in Sec. 3.3.

### 3.2.3 *Tip-tilt measurement performance*

The performance of the spot tracking algorithm is demonstrated in Fig. 10, where the tilt induced by the DM is highly comparable to that measured by the event tracking. The phase of the displacement oscillation induced on the DM does appear in this figure to go out of phase with the tracker's measurements at times; however, that could be attributed to observed inconsistencies in the DM induced tip-tilt due to signal processing issues. The laboratory computer used to control the DM had difficulties with processing speed while it transmitted the oscillating signal, resulting in an imperfect sinusoidal signal reaching the actuators. The tracker's performance was also confirmed empirically with the entire event stream and was observed to be in phase throughout.

The errors in measured tilt amplitude for each testing scenario were calculated to determine if similar levels of measurement accuracy could be achieved with several illumination conditions.

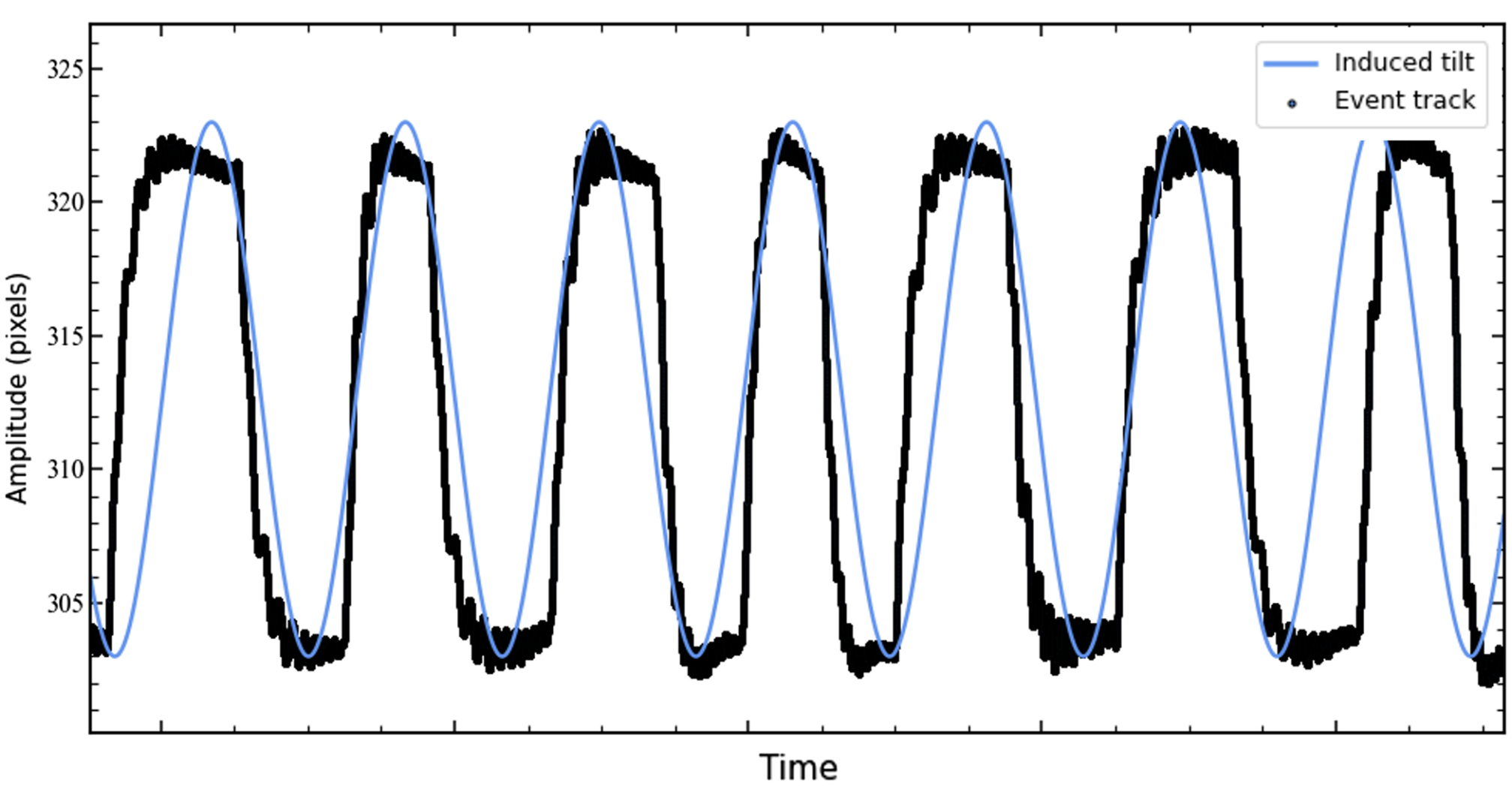

**Fig. 10** Comparison of the output of the event-tracking algorithm with the known DM-induced tilt. The blue line plots the signal sent to the DM rather than the true tilt induced by the DM.

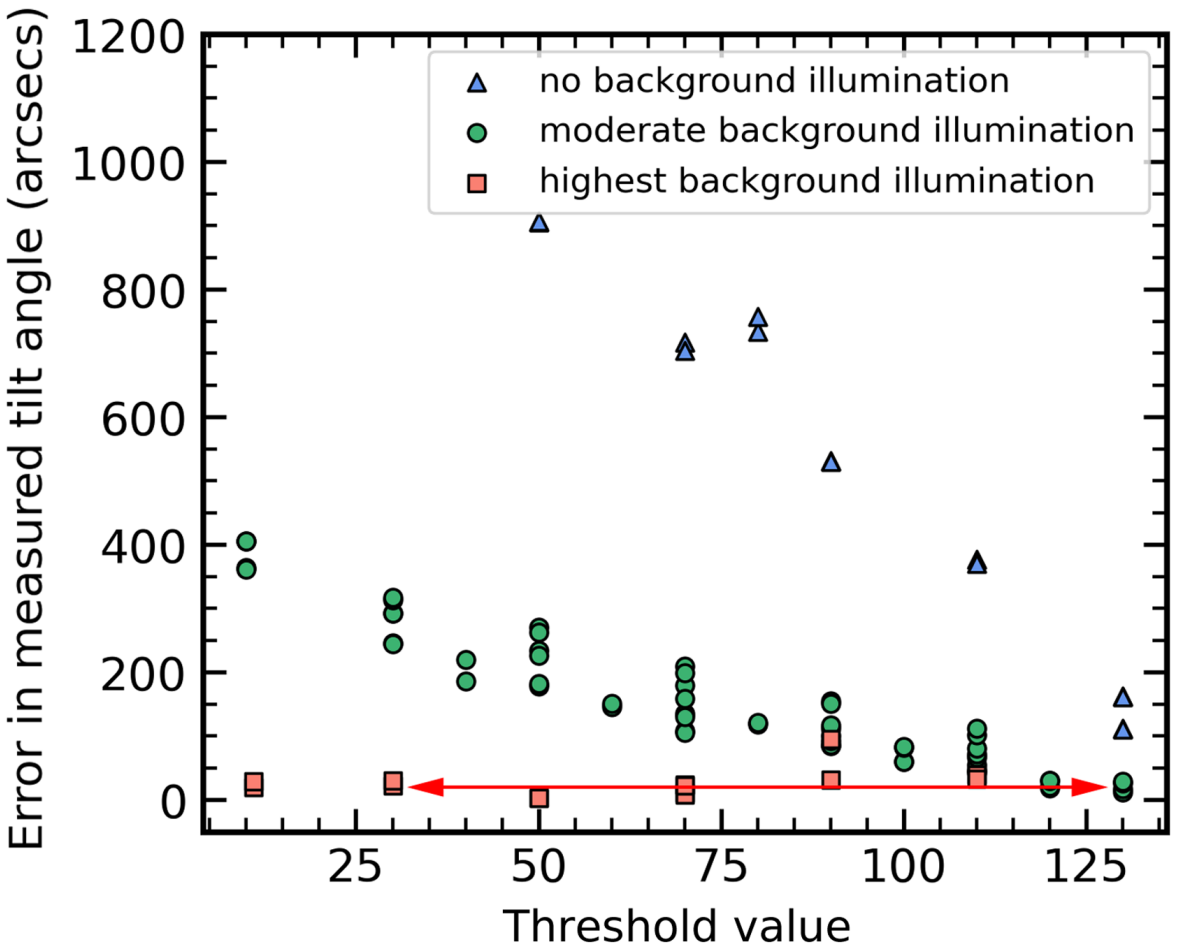

**Fig. 11** Error in measured tilt from the set tilt angle (true value) during a run of measurements, showing how an equivalent level of measurement accuracy can be achieved across multiple background illumination levels. More than one combination results in matching conditions, one of which is demonstrated by the red horizontal arrows between different conditions of background illumination. Different backgrounds were introduced using external light sources, and the threshold value of the event-based detector was set to a range of values. No background illumination is a dark room, moderate illumination is overhead ambient lighting, and the highest background illumination includes the addition of a torch adding light to the detector.[33]

Figure 11 demonstrates how the same three conditions and corresponding threshold values (highest background at threshold 30, moderate background at threshold 100, and no background at threshold 130) as in Fig. 7 are also matched in measurement error. This verifies that the conditions from different background illuminations can be made equivalent by the event-based detector by setting an appropriate threshold value.

In addition, Fig. 11 clearly demonstrates the event camera performing with higher accuracy when there is some level of background illumination. The highest background level produces measurement errors of an average of 1.5″, which is 0.5% of the tilt amplitude. This also agrees with previous studies[29] demonstrating that shot noise dominates at lower photocurrents and reduces to be significantly close to zero at higher overall illumination levels. In different illumination conditions, this behavior could be taken advantage of in lab-based activities by adding an artificial background to the field of view to improve the tilt measurement error. In wavefront sensing applications, this characteristic would result in daytime operations being advantageous over nighttime. The event-processing algorithm in this work also was observed to produce more accurate results when the laser spot moving on the detector had minimal overlap of the spot between its outer locations (i.e., left and right extremes) during tilt movement and was reasonably uniform in shape, further making these higher background conditions more optimal for this particular event-based detector application. This unique behavior of event-based sensors under different background conditions could thus be taken advantage of in whichever way suits each application.

To further verify this behavior of the sensor under high background illumination, measurements of tilt amplitudes near the camera plate scale were taken under the lower error conditions of high background illumination. Figure 12 demonstrates that each of the measurements could be made within an error of 1 pixel or 15″. This confirms that, with adjustments to the conditions such as background illumination level and event threshold values, the tilt measurements can be made with high accuracy. This is essential for implementing the time-delay method, where small tilt measurements are required.

The tilt wavefront error was calculated using Eq. (9)[34]

$$\Delta\omega_{\text{tilt}} = 2a\frac{y}{f} = n\lambda, \tag{9}$$

$$n = 2a\frac{y}{f}\frac{1}{\lambda},$$

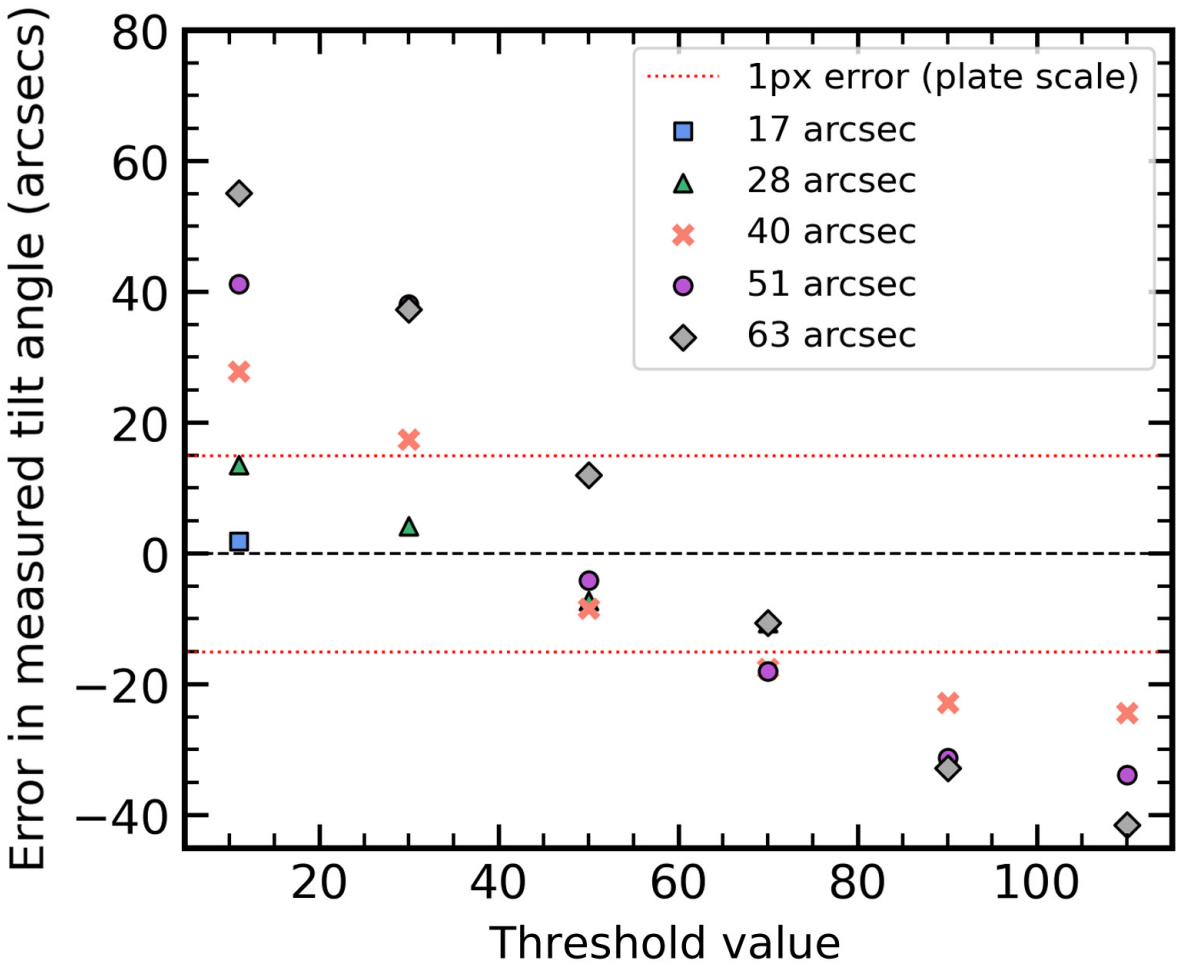

**Fig. 12** Tilt measurements taken close to the plate scale of 15″ under high background illumination levels. The red dotted line indicates where the error is 1 pixel or 15″.

where $a$ is the circular aperture radius, $f$ is the focal length of the camera, and $n$ is the number of waves of tilt introduced across the pupil diameter for a lateral displacement of $y$. The experimental results in Fig. 11 for the best measurement conditions produce a value of $n = 0.339$, which is an error in tilt measurement of approximately $\frac{\lambda}{3}$.

### 3.3 Discussion

As demonstrated, the behavior of an event-based sensor changes depending on the level of background illumination. Ideally, the background illumination will have a higher illumination relative to the tip-tilt signal level to minimize extra events being triggered due to shot noise while still minimizing noise around the signal source from leak events. For on-sky applications, this is advantageous for daytime wavefront sensing. Appropriate adjusting of the event-based detector's threshold values is fundamental to minimize this error term. The two noise sources can be balanced to optimize the noise output of the event-based detector. This is a significant benefit to tip-tilt sensing in a wider range of conditions, including daytime operations for free-space optical communications.

The average position tracker algorithm was able to update the tilt measurements with each event, making full use of the high-speed measurement capabilities of microsecond temporal resolution. This enabled the measurement of tilt down to an error of $\approx 1$ pixel once the event threshold values were appropriately chosen. For LGS tip-tilt sensing, this level of accuracy is beneficial for measuring small-scale differential tip-tilt, as required in the time-delay method.

The error contributions of the event-based sensor in this application include some clear differences from that of a traditional frame-based sensor. The accumulated error due to the integration time is not a factor in this operating principle as the operating principle of event-based detectors does not involve capturing frames. Although this theoretically would increase the sky coverage of an event-based wavefront sensor, it also needs to be considered whether the brightness changes are significant enough to trigger sufficient events (this item will be studied in future work). In addition, the lower data bandwidth of an event stream increases the speed at which real-time wavefront calculations can be made. However, small movements require extra consideration when locating the center of gravity of a spot on the sensor. Where a classical detector shows the whole spot area at all times, an event-based detector only displays the edges of the spot as the central area might not be changing brightness. This can contribute to the overall error of measured tip-tilt if the spot-tracking algorithm is not designed to properly handle this output. This could limit how accurately small tip-tilt movements could be measured using an event-based sensor and requires further exploration.

This research is a proof-of-concept to validate the event-based sensor for tip-tilt sensing. Although the accuracy required for diffraction-limited correction of atmospheric tip-tilt has yet to be tested, the potential of the event-based wavefront sensor for this application remains undisputed. To verify this capability, future work could include on-sky measurements of absolute tip-tilt using an NGS source, aiming to demonstrate the performance of the measurement accuracy below the diffraction limit. This will be undertaken using the Planewave RC700 telescope in the ANU Quantum Optical Ground Station; a 70-cm telescope. The next steps will focus on demonstrating measurement accuracy within an order of magnitude of the diffraction limit, serving as a proof of concept for the sensitivity of event-based sensors in detecting differential tip-tilt. This will be tested using both the full telescope aperture and a section of it to measure the sensor's limitation at different diffraction limit magnitudes. Further work will involve exploring new techniques such as introducing known illumination changes into the event-based sensor to increase the rate of triggered events.

## 4 Conclusion

The tip-tilt retrieval problem arises out of the need to measure tip-tilt on an LGS. The time-delay method is a proposed approach to address this challenge. However, the differential tip-tilt (that results from the time delay between upward and downward propagation of an LGS) would cause LGS spot movements with very small magnitudes, which are difficult to detect. The results in this paper indicate that the event-based sensor could be used to detect such small movements due to its increased sensitivity and lower noise levels. Future work in this area will include modifications to the experimental setup to reduce the plate scale and confirm the detector's capability for measuring tip-tilt on the milliarcsecond scale to further the investigation of its application to the time-delay method.

Based on this research, event-based detectors pose an advantage for daytime wavefront sensing due to their dynamic range. The experiments conducted demonstrated how increasing the overall illumination (daytime sky background) enables the detection of a tip-tilt source with higher accuracy and more consistency. The event-based detector's immunity to a constant sky background allows it to sense small movements despite noise levels. This is of particular use for laser communications applications that make use of AO in the daytime. The characterization of the detector outlined in this report demonstrates how this can be achieved through the appropriate choice of event threshold values.

Event-based detectors were tested to demonstrate the effect of key characteristics in the application of tip-tilt retrieval. A test bench was set up with a CenturyArks event-based detector. The detector's logarithmic response to changes in illumination has been characterized, which allows for a more informed choice when tuning the event threshold values and compensation of noise sources. This enables the optimization of the application of event-based detectors to different wavefront sensing conditions, such as that of astronomy or daytime free-space optical communications, with a comparative accuracy in tip-tilt measurement. The outcomes of this work suggest that the event-based sensor might be a strong contender for retrieving atmospheric tip-tilt information using LGSs and the time-delay method.

### Code and Data Availability

Data and code underlying the results presented in this paper are not publicly available at this time but may be obtained from the authors upon reasonable request.

### Acknowledgments

The authors would like to thank Esteban Vera and Vicente Westerhout from Pontificia Universidad Católica de Valparaíso for the collaboration and very productive discussions about their experiences with event-based detectors. This material is based on a work supported by the Air Force Office of Scientific Research under Award No. FA2386-23-1-4006.

**Monique Cockram** is a research assistant in the Laser Guide Star Adaptive Optics (LGS-AO) group at the Australian National University Advanced Instrumentation and Technology Centre. Her MSc thesis was working on the tip-tilt retrieval problem for LGS and laser communications. She currently leads the work on event-based detectors for wavefront sensing.

**Noelia Martinez Rey** is the instrument scientist lead of the Laser Guide Star Adaptive Optics (LGS-AO) group at the Australian National University Advanced Instrumentation and Technology Centre. She leads Australia's contribution to the next-generation AO facility for the Subaru Telescope and the development of the LGS facility for the Giant Magellan Telescope. Besides astronomy, she is a PI and Co-PI of several grants involving laser guide stars and adaptive optics technology translation to laser communications.